\pdfoutput=1
\documentclass[preprint,12pt]{elsarticle}

\usepackage[T1]{fontenc}
\usepackage{amsmath,amssymb,bm}
\usepackage{booktabs}
\usepackage[hidelinks]{hyperref}
\usepackage{graphicx}
\usepackage{caption}

\journal{Nuclear Physics A}
\biboptions{sort&compress}

\newcommand{\Wp}{W_{\mathrm{p}}}
\newcommand{\ReWp}{\operatorname{Re}\Wp}
\newcommand{\ImWp}{\operatorname{Im}\Wp}
\newcommand{\MeV}{\mathrm{MeV}}

\providecommand{\Wp}{W_{\mathrm{p}}}
\providecommand{\ReWp}{\operatorname{Re}\Wp}
\providecommand{\ImWp}{\operatorname{Im}\Wp}
\providecommand{\MeV}{\mathrm{MeV}}

\begin{document}

\begin{frontmatter}

\title{Non-uniqueness of the amplitude reconstruction in the inverse scattering problem and a refinement of  H\"ohler's resonance clustering}

\author[rbi,tesla]{Alfred \v{S}varc\corref{cor1}}
\ead{alfred.svarc@gmail.com}
\cortext[cor1]{Corresponding author.}

\affiliation[rbi]{organization={Ru{\dj}er Bo\v{s}kovi\'c Institute},
  addressline={Bijeni\v{c}ka cesta 54},
  city={Zagreb},
  postcode={10000},
  country={Croatia}}

\affiliation[tesla]{organization={Tesla Biotech d.o.o.},
  addressline={Mandlova 7},
  city={Zagreb},
  postcode={10000},
  country={Croatia}}
  
\begin{abstract}
  Reaction-amplitude reconstruction in the inverse scattering problem is
limited by the bilinear form of all reaction observables. At fixed
kinematics, $N$ complex amplitudes contain $2N$ real quantities, whereas
even complete measurements determine at most $2N-1$ parameters. The uniqueness of the inverse scattering problem 
therefore depends on whether additional physical constraints (like unitarity, for instance) supply
the missing information. We examine the consequences for resonance
identification, focusing on angular-momentum assignments.

Building on published truncation results that demonstrate that pole-bearing contributions of the infinite partial wave decomposition are redistributed due to fitting effective partial waves, we identify further mechanisms that move the positions of  pole-bearing terms  among effective partial waves as exact
and practical ambiguities associated with imposing coupled-channel
unitarity. An explicit $4\times4$ counterexample shows that different
on-shell unitary amplitudes can produce identical measurable bilinears.
Practical unitary descriptions can also yield different pole content
while remaining compatible with experimental uncertainties.

We update H\"ohler's clustering using the thirteen four-star $N^\ast$
resonances in PDG 2026 and the Bonn--Gatchina results published in 2025.
Nearby poles with different partial-wave assignments occur for the PDG compilation, but also within the
Bonn–Gatchina framework,  so their proximity is not solely a consequence
of the PDG compilation procedure. Some groupings persist, while others
change appreciably. 

We discuss a conditional interpretation in which
truncation redistributes a common pole-bearing contribution among
effective partial waves, with additional reconstruction uncertainties
potentially accounting for the departure from coincident pole positions.
These considerations allow, but do not establish, a common origin for
the members of a cluster. The necessity of separate states and the
stability of their angular-momentum assignments remain questions for
further investigation.

\end{abstract}
\begin{keyword}
Inverse scattering problem; reaction-amplitude reconstruction; continuum ambiguity; partial-wave analysis; resonance poles; angular-momentum ambiguity; nucleon resonances; Höhler clustering
\end{keyword}

\end{frontmatter}



\section{Introduction}
\label{sec:introduction}

Scattering theory describes a reaction through a set of complex amplitudes constrained by symmetries, analyticity, and unitarity. Experimental measurements, however, provide observables rather than amplitudes. Determining which properties of the amplitudes can be inferred uniquely from those measurements is the central question of the inverse scattering problem. This distinction is essential for resonance spectroscopy: pole positions, residues, and angular-momentum assignments are obtained through amplitude reconstruction and analytic continuation, rather than being measured directly.

At a fixed kinematic point, \(N\) complex reaction amplitudes comprise \(2N\) real quantities. According to the Born rule, all reaction observables are determined by bilinear combinations of these amplitudes and can generically determine at most \(2N-1\) independent real quantities~\citep{Barker1975,Chiang1997}. One real parameter, therefore, remains undetermined by the measurements. Additional physical requirements, such as analyticity and unitarity, introduce relations among the amplitude components. The question is whether these relations uniquely fix the remaining parameter in terms of the experimentally determined quantities, turning the 2N-real parameter problem into a 2N-1 constrained one. Merely reducing the number of independent parameters is insufficient if several admissible solutions remain. And we show that they indeed do.  Across the kinematic domain, the undetermined parameter can vary with energy and angle, and its angular dependence can change the partial-wave decomposition.

A central issue of our paper is the focus on the angular momentum assignment to an extracted resonance. In our earlier single-channel studies~\citep{Wunderlich2017,Svarc2018}, we explicitly connected continuum ambiguity to uncertainty in resonance quantum-number identification. An angle-dependent common phase transformation can redistribute pole-bearing contributions among partial waves while leaving all reaction observables unchanged. A resonance contribution can therefore remain present, while its association with a particular angular-momentum sector depends on the reconstructed amplitude. Our principal concern is this uncertainty in angular-momentum assignment. Angular momentum remains a natural classification of physical states; but what is questioned is the unique experimental identification of that classification. Here, orbital angular momentum \(L\) is distinguished from total angular momentum \(J\); the nucleon resonances considered below are classified by \(J^P\).

Amplitude non-uniqueness in inverse scattering problem has been investigated for more than five decades. Earlier studies established uniqueness results under specified conditions but also identified nontrivial ambiguities compatible with imposed unitarity constraints~\citep{Atkinson1972,Bart1973}. Such phase-ambiguous solutions remain a subject of investigation~\citep{Dersy2024}. The existence of amplitude ambiguity is therefore established background for the present work.

We examine its consequences for the interpretation of reconstructed resonance spectra, distinguishing two mechanisms with different roles: partial-wave truncation and ambiguity in unitarization.

The truncation mechanism was investigated in Ref.~\citep{Svarc2026}. When bilinear observables are fitted with a truncated partial-wave expansion, the extracted coefficients can contain pole-bearing contributions from different underlying angular-momentum sectors. In the idealized redistribution mechanism, the same analytic singularity appears in several extracted partial waves without changing its position. It therefore produces coincident pole images carrying different partial-wave labels. This provides a possible explanation for the common origin of poles assigned to different angular momenta. It does not, by itself, explain why those poles would occupy nearby but distinct positions. This could be explained by amplitude non-uniqueness in inverse scattering theory. 

One reason is certainly the incompleteness of the database. However, there is another, more fundamental reason.  Amplitude non-uniqueness has been investigated for more than five decades. The source of non-uniqueness is the way how unitarity is imposed.  Earlier studies established uniqueness results, but only under specific conditions, and also identified nontrivial ambiguities compatible with differently imposed unitarity constraints~\citep{Atkinson1972,Bart1973}. The existence of amplitude ambiguity is therefore established background for the present work. Above the first inelastic threshold, unitarity must be imposed on the coupled-channel system. We distinguish two forms of ambiguity in reconstructing amplitudes under this requirement. The first is exact ambiguity: different unitary amplitudes produce identical measurable bilinears. The four-channel construction presented in the Appendix demonstrates such non-uniqueness for the separately measurable channel bilinears of the example. The second is practical model dependence: different unitarization prescriptions produce different amplitudes and slightly different bilinears, but these differences can remain indistinguishable within experimental uncertainties. These prescriptions implement the same unitarity requirement through different choices of analytic input, channel content, subtraction terms, and background parametrizations~\citep{CDD1956,Cutkosky1979,Manley1992,Anisovich2012,Roenchen2022}.

Unlike the redistribution of a fixed pole due to the truncation effect upon angular momentum redistribution, alternative unitary reconstructions need not preserve the same analytic singularities. Different model implementations can consequently yield different pole positions and residues while describing the available measurements comparably well. Exact bilinear ambiguity also leaves the analytic amplitude undetermined, although the four-channel construction alone does not establish that its alternative solutions have different poles; that requires an energy-dependent analytic realization. The following section examines these two forms of unitarization ambiguity and distinguishes what they establish about amplitude reconstruction from their possible consequences for pole extraction.

These considerations motivate a refinement of H\"ohler's resonance clustering~\citep{Hoehler1983}. We update the pole pattern using the established four-star nucleon spectrum of PDG compilation~\citep{PDG2026} and of Bonn-Gatchina model~\cite{Sarantsev2025}. We consider a two-stage interpretation. Truncation-induced redistribution supplies the proposed common origin of pole contributions found in different partial waves. Subsequent ambiguities in their unitary reconstruction can contribute to departures from the ideal coincidence, along with uncertainties in the experimental input. Unitarization ambiguity is thus considered a possible source of the finite spread of a cluster, rather than the mechanism generating its common pole origin.

The observed clustering does not by itself prove that its members are reconstructed images of one physical resonance, nor does the present analysis quantify  the angular momentum of starting, true resonance. It identifies a possible connection between the pole pattern and the limitations of the inverse scattering problem. Within this interpretation, a cluster may suggest a common underlying structure without uniquely identifying its angular momentum. The central distinction is therefore between the robustness of a resonance signal and the robustness of the quantum-number assignment inferred from its reconstructed partial-wave content.

\section{Ambiguities of reaction amplitudes in inverse scattering}
\label{sec:amplitude-ambiguities}

We consider two mechanisms affecting the reconstruction of resonance content from bilinear observables: partial-wave truncation and ambiguities associated with imposing unitarity.

\subsection{Ambiguities due to truncation}
\label{subsec:truncation-ambiguities}

Truncation is an unavoidable feature of practical partial wave reconstruction since finite experimental information and computational resources do not permit an unrestricted determination of infinitely many partial waves.

The truncation mechanism developed in Ref.~\citep{Svarc2026} arises because measured bilinear observables contain contributions from partial waves omitted from the fitted amplitude, including their interference with retained waves. When these observables are fitted with a truncated expansion, the retained coefficients adjust to accommodate the effects of the omitted contributions. The reconstructed amplitude is therefore generally an effective description whose coefficients need not coincide with the corresponding coefficients of the exact amplitude. The most important fact to stress is that truncation does not mean compensating for the higher partial wave only; due to the mathematics, the coefficients of the lower partial waves are reshuffled too. 

This effect was demonstrated in a minimal scalar example in which bilinear quantities generated by an amplitude containing \(S\), \(P\), and \(D\) waves were fitted using only \(S\) and \(P\) waves. Although the fitted amplitude contains no explicit \(D\) wave, its reconstructed lower-wave coefficients depend on combinations involving the original \(D\) wave contribution. Truncation thus induces mixing in the extracted coefficients through the nonlinear reconstruction from bilinear observables.

Each fitted coefficient still multiplies a basis function with definite angular momentum. Its inferred dynamical content, however, can contain contributions originating from several underlying partial waves. Since those contributions may carry resonance poles, the angular momentum label of the fitted coefficient does not, by itself, uniquely identify the underlying angular momentum of a resonance contribution appearing in it. The extracted resonance content must consequently be interpreted as belonging to an effective, truncation-dependent reconstruction.

The present discussion deals with this mixing mechanism and does not include a separate analysis of analytic continuation\footnote{ For a given amplitude, analytic continuation along a specified admissible path is unique wherever it exists. Under suitable analyticity and boundary-regularity assumptions, exact complex boundary values throughout an open physical-energy interval determine the amplitude and its continuation. However, finitely many sampled values—even of the complex amplitude—do not generally ensure uniqueness without further constraints. Bilinear observables alone leave additional amplitude information undetermined. Specifying only pole and cut locations does not fix residues, cut discontinuities, or the analytic background; different reconstructions compatible with finite data can therefore have different continuations without introducing additional cuts. Experimental uncertainties can also make the reconstruction strongly unstable. These limitations of reconstructing the energy dependence from finite or inexact data should be distinguished from truncation of the angular expansion discussed here.}. For the clustering interpretation considered here, the relevant idealized limit is the redistribution of a common pole-bearing term without displacement of its singularity. Its images then occur at the same complex energy in different extracted partial waves. This provides a possible common origin for cross-wave pole groupings, while departures from exact coincidence require additional consideration. Preservation of pole positions in this idealized redistribution should not be understood as a general property of every truncated fit.

\subsubsection{Exact ambiguity}
\label{subsubsec:exact-unitarity-ambiguity}

Unitarity introduces relations among reaction amplitudes, but these relations need not determine uniquely the information missing from bilinear measurements. Exact ambiguities compatible with unitarity have been studied in the literature~\citep{Atkinson1972,Bart1973,Dersy2024}. Their existence means that non-uniqueness need not originate from experimental errors: different amplitudes can satisfy the imposed constraints and reproduce exactly the same measured observables.

For coupled-channel reconstruction, the relevant distinction is between bilinears measurable within an individual reaction and interference products involving different reactions. Separately measured channel observables do not generally supply the latter information. Consequently, even exact knowledge of all separately measurable channel bilinears need not determine a unique unitary matrix.

\ref{app:four-channel-ambiguity} provides another explicit four-channel example. A continuous parameter generates different symmetric, unitary matrices while leaving every reaction intensity unchanged, including those of the elastic channels. The alternatives are not ordinary rephasings of the channel states: their eigenvalues differ. Thus, the ambiguity survives even when all four channels are included and their separately measurable bilinears are known exactly. Neither omitted channels nor finite experimental precision are responsible for this non-uniqueness.

These examples do not establish that non-uniqueness occurs in every reaction or constitutes generic behavior. Their logical role is to provide counterexamples to the unconditional proposition that imposing unitarity uniquely determines reaction amplitudes from measurable bilinears. A single counterexample is sufficient to disprove that proposition. Uniqueness may still hold under additional assumptions or in particular physical settings, but it cannot be inferred from unitarity and bilinear information alone as a universal mathematical result.

The four-channel construction isolates this issue at fixed energy within one angular-momentum sector. It does not itself demonstrate mixing between different angular momenta or changes in complex-energy pole positions. Those consequences additionally depend on the angular and energy dependence of the alternative reconstructions.

\subsubsection{Practical model dependence}
\label{subsubsec:practical-unitarity-dependence}

A second ambiguity arises from the additional assumptions required to implement unitarity in practical amplitude reconstruction. Unitarity specifies relations among amplitudes but does not determine the complete dynamics. Different prescriptions introduce different analytic input, background parametrizations, subtraction terms, and treatments of explicit or effective channels. Examples include analytic multichannel parametrizations, \(K\)-matrix descriptions, and dynamical coupled-channel equations~\citep{CDD1956,Cutkosky1979,Manley1992,Anisovich2012,Roenchen2022}. These approaches implement the same physical requirement through different constructions.

Unlike the exact ambiguity discussed above, different constructions generally produce different amplitudes and different bilinears. Nevertheless, after fitting the same experimental data, their predicted observables can differ by amounts that the measurements cannot resolve. Finite statistical and systematic uncertainties, together with incomplete energy and angular coverage, can therefore allow several reconstructions to remain compatible with the available information. Selecting one prescription then selects an amplitude that is determined partly by its model assumptions.

The distinction from exact ambiguity is essential. In the exact case, alternative amplitudes reproduce mathematically identical measurable bilinears. In the practical case, their bilinears differ, but the experimental information does not discriminate sufficiently between them. This second ambiguity can remain even if the fitting procedure determines a unique solution within each chosen model. Uniqueness within a parametrization does not establish independence from the parametrization itself.

The resulting uncertainty extends to resonance extraction. Alternative reconstructions need not contain identical pole positions, residues, or pole counts, even when their predictions are comparably consistent with the measurements. These differences arise from the reconstructed analytic amplitudes; analytic continuation is unique once each amplitude is fixed. Moreover, small differences between observables on the physical axis do not, by themselves, guaranty equally small differences between extracted pole parameters.

Within the interpretation of clustering considered below, this model dependence has a different role from truncation-induced redistribution. The idealized truncation mechanism supplies the proposed common origin of coincident pole images in different partial waves. Different unitarization prescriptions can alter the reconstructed effective amplitudes and displace their extracted poles, providing a possible contribution to the finite spread of the observed clusters. Unitarization does not itself mix the exact angular-momentum sectors, and neither the size of these displacements nor the survival of a cluster follows from unitarity alone. The proposed explanation of the spread must therefore be assessed through comparisons of alternative reconstructions.
\section{Update of H\"ohler's resonance clustering}
\label{sec:hoehler-update}

We re-examine H\"ohler's clustering using the PDG 2026 pole estimates~\cite{PDG2026} and the
results of the Bonn-Gatchina 2025 analysis~\cite{Sarantsev2025} before comparing the two pole
distributions.

\subsection{PDG 2026 pole estimates}
\label{subsec:pdg2026}

We consider the thirteen four-star excited $N^\ast$ resonances in the
PDG 2026 compilation~\cite{PDG2026}. Isospin is therefore fixed to $I=1/2$;
$\Delta$ resonances and lower-star states are excluded. The numerical input
is given in Table~\ref{tab:pdg2026-poles}.

For each resonance, we write the pole position as $\Wp=M_{\mathrm p}-\frac{i}{2}\Gamma_{\mathrm p}$.

Figure~\ref{fig:pdg2026-clusters} displays the approximate PDG pole estimates
in the complex-energy plane, together with a
three-dimensional representation that retains the partial-wave labels.
The PDG tabulates $-2\operatorname{Im}\Wp$; both its central values and
interval limits are divided by two for the plotted ordinate. The line
segments represent quoted compilation ranges, not correlated statistical
uncertainties from a common fit. Positive- and negative-parity states are
shown by upward blue and downward red triangles, respectively.

The third coordinate is the partial-wave plotting index
\begin{equation}
  \eta_{\mathrm{PW}}=L+J+\frac12.
  \label{eq:eta-pw}
\end{equation}
For $\pi N$ scattering, $J=L\pm\frac12$, and this index assigns successive
integer levels to $S_{11},P_{11},P_{13},D_{13},D_{15},F_{15},\ldots$.
It distinguishes the complete $(L,J)$ partial-wave assignments, which
neither $L$ nor $J$ alone can do\footnote{The standard Dirac angular index
$\kappa=(L-J)(2J+1)$ equivalently labels the pair $(L,J)$ for spin-$1/2$
angular coupling~\cite{Szmytkowski2007}. Our plotting index is its positive
enumeration: $\eta_{\mathrm{PW}}=2\kappa$ for $\kappa>0$, and
$\eta_{\mathrm{PW}}=-2\kappa-1$ for $\kappa<0$.}.
The vertical ticks carry the full partial-wave names. Equal spacing specifies
their order, not a physical distance between states. Resonances assigned to
the same partial wave share one level and are distinguished by their pole
positions in the complex-energy plane.

\clearpage
\begin{table}[p]
\centering
\caption{PDG 2026 approximate pole estimates and quoted intervals used in
Fig.~\ref{fig:pdg2026-clusters}. All numerical entries are in MeV. The last
two columns contain \(-\ImWp=\Gamma_{\mathrm p}/2\), rather than the full pole
width \(-2\ImWp\) tabulated by the PDG.}
\label{tab:pdg2026-poles}
\small
\setlength{\tabcolsep}{4.5pt}
\renewcommand{\arraystretch}{1.15}
\begin{tabular}{@{}lccccc@{}}
\toprule
State
& \(J^P\)
& \(\ReWp\)
& range of \(\ReWp\)
& \(-\ImWp\)
& range of \(-\ImWp\) \\
\midrule
\(N(1440)\) & \(1/2^+\) & 1370 & 1360--1380 &  95   &  90--102.5 \\
\(N(1520)\) & \(3/2^-\) & 1510 & 1505--1515 &  55   &  52.5--60 \\
\(N(1535)\) & \(1/2^-\) & 1510 & 1500--1520 &  55   &  40--65 \\
\(N(1650)\) & \(1/2^-\) & 1665 & 1650--1680 &  67.5 &  50--85 \\
\(N(1675)\) & \(5/2^-\) & 1655 & 1650--1660 &  67.5 &  60--75 \\
\(N(1680)\) & \(5/2^+\) & 1670 & 1660--1680 &  60   &  55--67.5 \\
\(N(1710)\) & \(1/2^+\) & 1700 & 1650--1750 &  60   &  40--80 \\
\(N(1720)\) & \(3/2^+\) & 1680 & 1660--1710 & 100   &  75--150 \\
\(N(1895)\) & \(1/2^-\) & 1910 & 1890--1930 &  55   &  40--70 \\
\(N(1900)\) & \(3/2^+\) & 1920 & 1900--1940 &  65   &  45--80 \\
\(N(2190)\) & \(7/2^-\) & 2050 & 1950--2150 & 200   & 150--250 \\
\(N(2220)\) & \(9/2^+\) & 2150 & 2130--2200 & 200   & 180--240 \\
\(N(2250)\) & \(9/2^-\) & 2150 & 2100--2200 & 210   & 175--250 \\
\bottomrule
\end{tabular}
\end{table}
\clearpage

\begin{figure}[p]
\centering
\includegraphics[width=\textwidth]{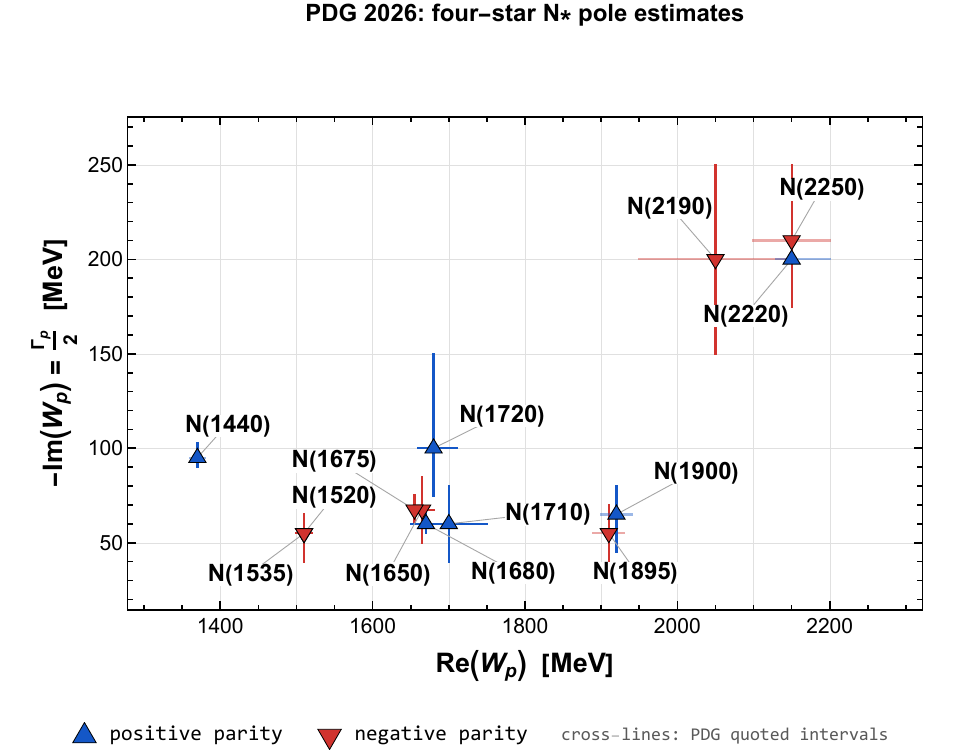}
\caption{Pole positions of the thirteen four-star excited $N^\ast$
resonances selected from the PDG 2026 compilation: (a) projection onto the
complex-energy plane. Upward blue and downward red triangles denote
positive and negative parity, respectively. Horizontal and vertical line
segments show the quoted PDG ranges for the real part and half-width.
They are not correlated confidence regions from a common fit.}
\label{fig:pdg2026-clusters}
\end{figure}
\clearpage
\begin{figure}[p]
\ContinuedFloat
\centering
\includegraphics[width=\textwidth]{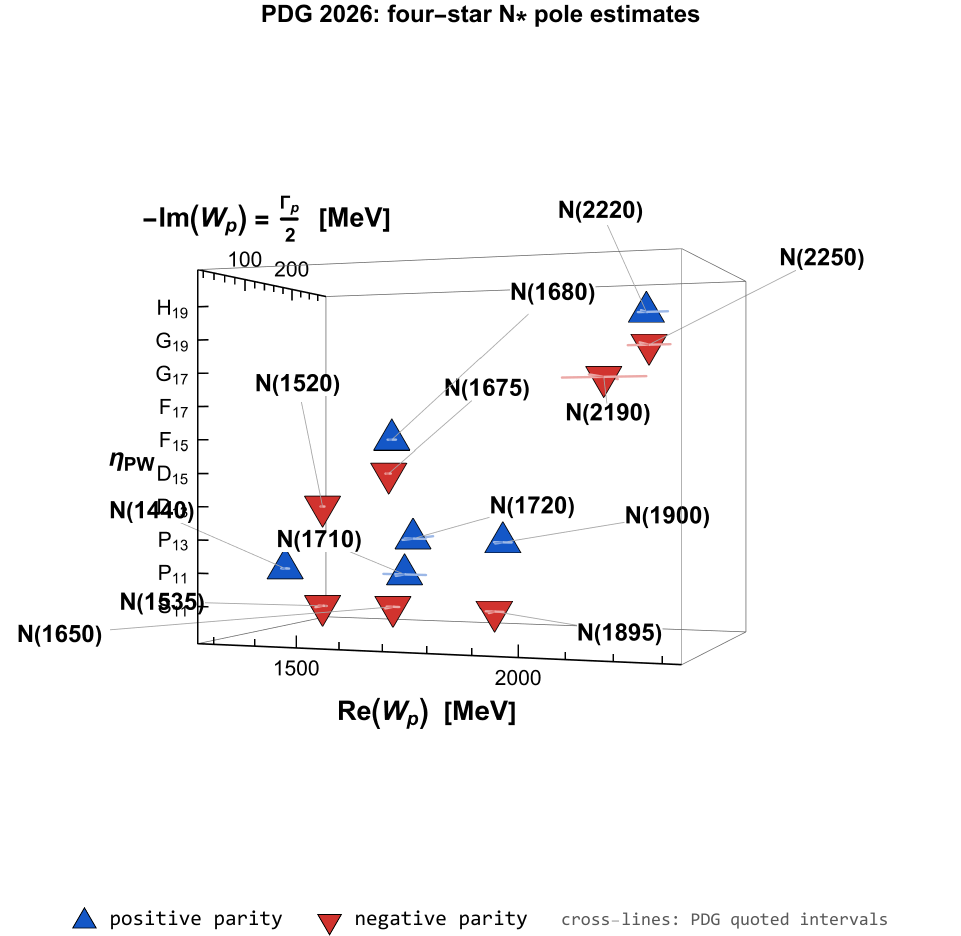}
\caption[]{Continued. (b) The same pole estimates displayed with
$\eta_{\mathrm{PW}}=L+J+\frac12$ as the third coordinate. The vertical ticks
carry full $\pi N$ partial-wave labels. Colours, symbols and quoted ranges
have the same meaning as in panel (a).}
\end{figure}
\clearpage
Four candidate groupings can be identified from the displayed central values.

The first contains $N(1520)$ in $D_{13}$ and $N(1535)$ in $S_{11}$.
Both estimates occur at
\begin{equation}
  \Wp=(1510-i\,55)\ \MeV.
\end{equation}
This coincidence concerns rounded PDG estimates and does not establish
exact degeneracy of the underlying poles.

The second contains $N(1650)$ in $S_{11}$, $N(1675)$ in $D_{15}$,
$N(1680)$ in $F_{15}$, and $N(1710)$ in $P_{11}$, with
\begin{equation}
\begin{aligned}
  \Wp[N(1650)]&=(1665-i\,67.5)\ \MeV,\\
  \Wp[N(1675)]&=(1655-i\,67.5)\ \MeV,\\
  \Wp[N(1680)]&=(1670-i\,60)\ \MeV,\\
  \Wp[N(1710)]&=(1700-i\,60)\ \MeV.
\end{aligned}
\end{equation}
These four central positions span $45\ \MeV$ in the real part and
$7.5\ \MeV$ in the half-width, while their assigned orbital angular momenta
cover $L=0,1,2,3$.

The third consists of $N(1895)$ in $S_{11}$ and $N(1900)$ in $P_{13}$:
\begin{equation}
\begin{aligned}
  \Wp[N(1895)]&=(1910-i\,55)\ \MeV,\\
  \Wp[N(1900)]&=(1920-i\,65)\ \MeV.
\end{aligned}
\end{equation}
Their central positions differ by $10\ \MeV$ in each plotted coordinate.

The fourth contains $N(2220)$ in $H_{19}$ and $N(2250)$ in $G_{19}$:
\begin{equation}
\begin{aligned}
  \Wp[N(2220)]&=(2150-i\,200)\ \MeV,\\
  \Wp[N(2250)]&=(2150-i\,210)\ \MeV.
\end{aligned}
\end{equation}
Their central pole masses coincide, and their half-widths differ by
$10\ \MeV$. Although both have $J=9/2$, their partial waves have different
orbital angular momenta and opposite parities.

The remaining three states are not included in these groupings. $N(1440)$
is isolated. $N(1720)$ has a central pole mass near the second group but a
larger half-width. The broad quoted range of $N(2190)$ approaches the
high-mass pair, although its central pole mass lies about $100\ \MeV$ below
that pair.

Here, the term \emph{cluster} describes proximity of the displayed central
values; no statistical clustering significance is assigned. The concentration
of poles from different partial waves provides a modern counterpart of
H\"ohler's observation. The figure does not, by itself, demonstrate that the
members of a cluster originate from one physical resonance.
\subsection{Bonn--Gatchina 2025 pole results}
\label{subsec:bnga2025}

To examine the pole distribution within a single analysis framework, we use
the Bonn--Gatchina results published in 2025 by Sarantsev
et al.~\cite{Sarantsev2025}. The solution is denoted BnGa-2024 in that
publication. We retain the same thirteen four-star excited $N^\ast$
resonances as in the preceding subsection, with $I=1/2$. All pole values
are taken from Table~IV of Ref.~\cite{Sarantsev2025}; the numerical input
is listed in Table~\ref{tab:bnga2025-poles}.

Figure~\ref{fig:bnga2025-clusters} shows the pole positions using the same
coordinates, parity symbols and partial-wave plotting index
$\eta_{\mathrm{PW}}$ defined in Eq.~\eqref{eq:eta-pw}. The mass and full
width quoted in the source are converted to
$\Wp=M_{\mathrm p}-i\Gamma_{\mathrm p}/2$. Both the full widths and their
uncertainties are divided by two for the ordinate
$-\operatorname{Im}\Wp$. The quoted uncertainties are estimated from the
spread of results under variations of the fit, including changes of the
model space and data treatment. The plotted bars are therefore not assigned
a common statistical confidence level, and no correlations are inferred
from them.

\clearpage
\begin{table}[p]
\centering
\caption{Bonn--Gatchina pole values and quoted uncertainties for the thirteen
selected four-star $N^\ast$ resonances, taken from Table~IV of
Ref.~\cite{Sarantsev2025}. All numerical entries are in MeV. The two
uncertainty columns give the authors' quoted uncertainties for the real
part and half-width, respectively. Both the full pole widths and their
uncertainties have been divided by two.}
\label{tab:bnga2025-poles}
\small
\setlength{\tabcolsep}{4.5pt}
\renewcommand{\arraystretch}{1.15}
\begin{tabular}{@{}lccccc@{}}
\toprule
State & \(J^P\) & \(\ReWp\) & uncertainty
& \(-\ImWp\) & uncertainty \\
\midrule
\(N(1440)\) & \(1/2^{+}\) & 1366 & 3 & 96 & 2 \\
\(N(1520)\) & \(3/2^{-}\) & 1506 & 2 & 56 & 1.5 \\
\(N(1535)\) & \(1/2^{-}\) & 1494 & 7 & 57.5 & 5 \\
\(N(1650)\) & \(1/2^{-}\) & 1664 & 10 & 52.5 & 3 \\
\(N(1675)\) & \(5/2^{-}\) & 1662 & 5 & 65 & 4 \\
\(N(1680)\) & \(5/2^{+}\) & 1676 & 6 & 58.5 & 2 \\
\(N(1710)\) & \(1/2^{+}\) & 1696 & 10 & 77.5 & 7.5 \\
\(N(1720)\) & \(3/2^{+}\) & 1710 & 20 & 90 & 15 \\
\(N(1895)\) & \(1/2^{-}\) & 1905 & 15 & 92.5 & 12.5 \\
\(N(1900)\) & \(3/2^{+}\) & 1925 & 25 & 135 & 15 \\
\(N(2190)\) & \(7/2^{-}\) & 2130 & 35 & 185 & 22.5 \\
\(N(2220)\) & \(9/2^{+}\) & 2165 & 30 & 220 & 20 \\
\(N(2250)\) & \(9/2^{-}\) & 2220 & 40 & 235 & 20 \\

\bottomrule
\end{tabular}
\end{table}
\clearpage

\begin{figure}[p]
\centering
\includegraphics[width=\textwidth]{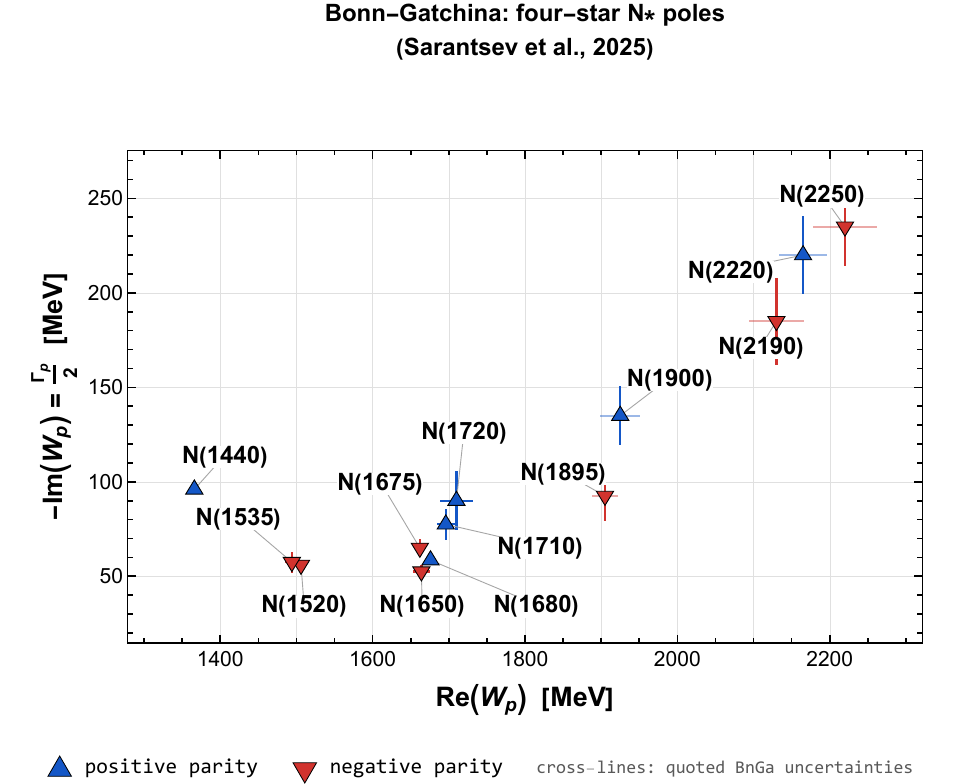}
\caption{Pole positions of the thirteen selected four-star excited
$N^\ast$ resonances from the Bonn--Gatchina analysis published in
2025~\cite{Sarantsev2025}: (a) projection onto the complex-energy plane.
Upward blue and downward red triangles denote positive and negative
parity, respectively. Horizontal and vertical line segments show the
quoted uncertainties in the pole mass and half-width.}
\label{fig:bnga2025-clusters}
\end{figure}
\clearpage

\begin{figure}[p]
\ContinuedFloat
\centering
\includegraphics[width=\textwidth]{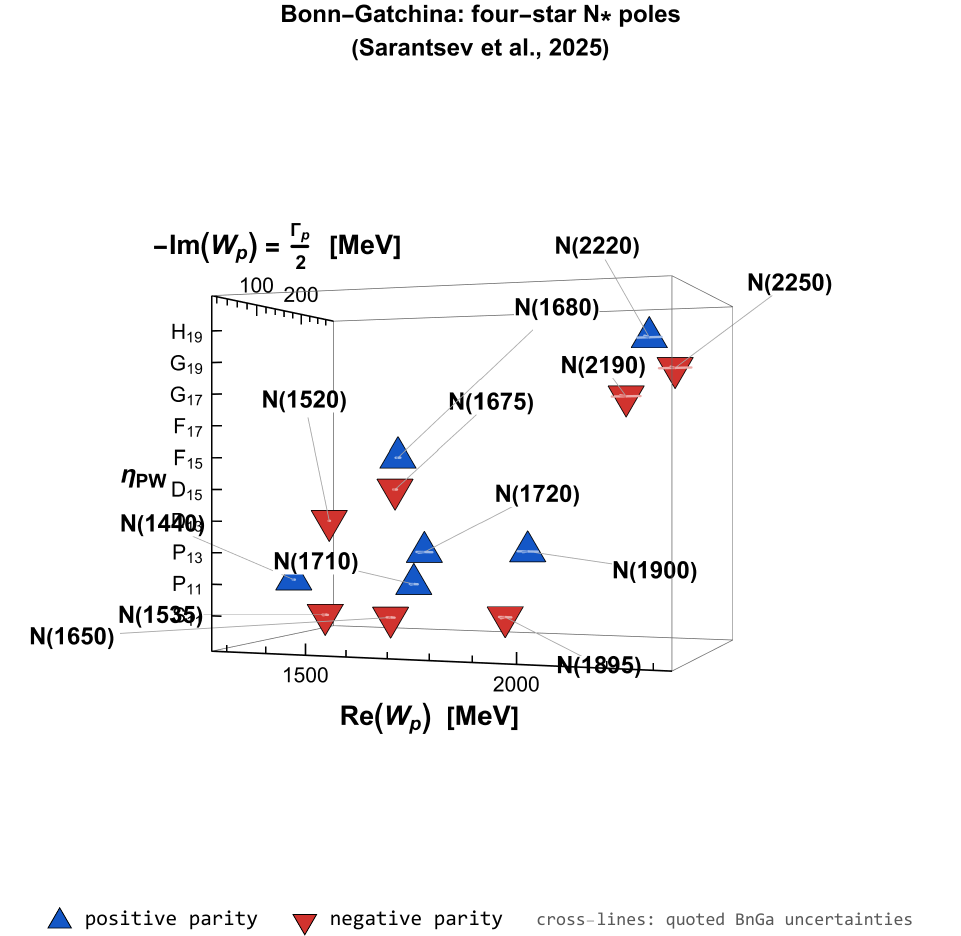}
\caption[]{Continued. (b) The same Bonn--Gatchina poles displayed with
$\eta_{\mathrm{PW}}=L+J+\frac12$ as the third coordinate and full
$\pi N$ partial-wave labels at the vertical ticks. Colours, symbols
and uncertainty bars have the same meaning as in panel (a).}
\end{figure}
\clearpage

For a consistent examination, we inspect the same four candidate
memberships identified in the PDG subsection. They are not the result
of a new statistical clustering procedure applied to the Bonn--Gatchina
values.

The first pair contains $N(1520)$ in $D_{13}$ and $N(1535)$ in $S_{11}$:
\begin{equation}
\begin{aligned}
  \Wp[N(1520)]&=(1506-i\,56)\ \MeV,\\
  \Wp[N(1535)]&=(1494-i\,57.5)\ \MeV.
\end{aligned}
\end{equation}
Their central pole masses differ by $12\ \MeV$, while their half-widths
differ by $1.5\ \MeV$.

The second group contains $N(1650)$ in $S_{11}$, $N(1675)$ in $D_{15}$,
$N(1680)$ in $F_{15}$, and $N(1710)$ in $P_{11}$:
\begin{equation}
\begin{aligned}
  \Wp[N(1650)]&=(1664-i\,52.5)\ \MeV,\\
  \Wp[N(1675)]&=(1662-i\,65)\ \MeV,\\
  \Wp[N(1680)]&=(1676-i\,58.5)\ \MeV,\\
  \Wp[N(1710)]&=(1696-i\,77.5)\ \MeV.
\end{aligned}
\end{equation}
These central positions span $34\ \MeV$ in the real part and $25\ \MeV$
in the half-width. Their assigned orbital angular momenta again cover
$L=0,1,2,3$.

For the third candidate pair, $N(1895)$ in $S_{11}$ and $N(1900)$
in $P_{13}$, the central values are
\begin{equation}
\begin{aligned}
  \Wp[N(1895)]&=(1905-i\,92.5)\ \MeV,\\
  \Wp[N(1900)]&=(1925-i\,135)\ \MeV.
\end{aligned}
\end{equation}
Their pole masses differ by $20\ \MeV$, and their half-widths by
$42.5\ \MeV$. This pair is appreciably separated in the imaginary
coordinate and does not have the same compact appearance as the first
pair.

The fourth candidate grouping contains $N(2220)$ in $H_{19}$ and
$N(2250)$ in $G_{19}$:
\begin{equation}
\begin{aligned}
  \Wp[N(2220)]&=(2165-i\,220)\ \MeV,\\
  \Wp[N(2250)]&=(2220-i\,235)\ \MeV.
\end{aligned}
\end{equation}
Their central pole masses differ by $55\ \MeV$, while their half-widths
differ by $15\ \MeV$. They share $J=9/2$ but have different orbital
angular momenta and opposite parities.

The three remaining selected states have central poles
\begin{equation}
\begin{aligned}
  \Wp[N(1440)]&=(1366-i\,96)\ \MeV,\\
  \Wp[N(1720)]&=(1710-i\,90)\ \MeV,\\
  \Wp[N(2190)]&=(2130-i\,185)\ \MeV.
\end{aligned}
\end{equation}
The first is isolated, whereas $N(1720)$ and $N(2190)$ lie near the
second and fourth candidate regions, respectively. The retained memberships
therefore do not define unique or exhaustive cluster boundaries.

The figure shows proximity of poles assigned to different partial waves
within one analysis framework, without requiring an average over results
from different analysis groups. The detailed comparison with the PDG
pattern is deferred to the next subsection. The figure does not, by
itself, demonstrate that the members of a cluster originate from one
physical resonance.

\subsection{Comparison of PDG 2026 and Bonn--Gatchina 2025}
\label{subsec:pdg-bnga-comparison}

The two preceding subsections display the same thirteen four-star excited
$N^\ast$ resonances, using a common pole convention and partial-wave
coordinate. Figures~\ref{fig:pdg2026-clusters}
and~\ref{fig:bnga2025-clusters} therefore permit a direct comparison of
the reported central pole positions. The PDG estimates summarize results
from different analyses, whereas the Bonn--Gatchina values are obtained
within one coupled-channel analysis framework.

For the same four candidate memberships, we characterize the extent of
each group $C$ by
\begin{equation}
\begin{aligned}
 R_M(C)&=\max_{r\in C}M_{{\mathrm p},r}
        -\min_{r\in C}M_{{\mathrm p},r},\\
 R_H(C)&=\max_{r\in C}\frac{\Gamma_{{\mathrm p},r}}{2}
        -\min_{r\in C}\frac{\Gamma_{{\mathrm p},r}}{2}.
\end{aligned}
\label{eq:cluster-comparison-spans}
\end{equation}
These quantities measure the spans of the central positions along the
mass and half-width coordinates. The quoted bars in the two sources have
different constructions, so we assign no statistical significance to
these descriptive spans.

\begin{table}[t]
\centering
\caption{Spans of the central pole positions for the same candidate
memberships in the PDG 2026 and Bonn--Gatchina 2025 results.
$R_M$ and $R_H$ are defined in Eq.~\eqref{eq:cluster-comparison-spans};
all numerical entries are in MeV. The memberships are retained for
comparison and are not selected by a statistical clustering algorithm.}
\label{tab:pdg-bnga-comparison}
\small
\setlength{\tabcolsep}{4pt}
\renewcommand{\arraystretch}{1.2}
\begin{tabular}{@{}p{0.40\textwidth}cccc@{}}
\toprule
Candidate members & \multicolumn{2}{c}{PDG 2026}
& \multicolumn{2}{c}{Bonn--Gatchina 2025} \\
\cmidrule(lr){2-3}\cmidrule(lr){4-5}
 & \(R_M\) & \(R_H\) & \(R_M\) & \(R_H\) \\
\midrule
\(N(1520),N(1535)\) & 0 & 0 & 12 & 1.5 \\
\(N(1650),N(1675)\)\newline \(N(1680),N(1710)\) & 45 & 7.5 & 34 & 25 \\
\(N(1895),N(1900)\) & 10 & 10 & 20 & 42.5 \\
\(N(2220),N(2250)\) & 0 & 10 & 55 & 15 \\
\bottomrule
\end{tabular}
\end{table}

The $N(1520)$--$N(1535)$ pair remains close in both representations.
The rounded PDG estimates coincide, while the Bonn--Gatchina central
positions differ by $12\ \MeV$ in mass and only $1.5\ \MeV$ in
half-width. The quartet comprising $N(1650)$, $N(1675)$, $N(1680)$
and $N(1710)$ also occupies a restricted region in both distributions.
Its mass span is smaller in Bonn--Gatchina, $34\ \MeV$ compared with
$45\ \MeV$, but its half-width span increases from $7.5$ to
$25\ \MeV$. Thus, proximity of these poles with different assigned
partial waves persists, although the shape of the group changes.

The correspondence is less pronounced for the other two memberships.
For $N(1895)$ and $N(1900)$, the half-width separation increases from
$10\ \MeV$ in the PDG estimates to $42.5\ \MeV$ in Bonn--Gatchina.
This pair therefore cannot be described as an equally compact cluster
in both representations. Its location also changes: the two central
half-widths increase from $55$ and $65\ \MeV$ to $92.5$ and
$135\ \MeV$, respectively. For $N(2220)$ and $N(2250)$, the common
central PDG pole mass of $2150\ \MeV$ is replaced by masses of
$2165$ and $2220\ \MeV$. Their mass separation is therefore
$55\ \MeV$, while the half-width separation changes from $10$ to
$15\ \MeV$.
Neither the rounded coincidences nor the detailed extent of every
candidate group are preserved.

The proximity of $N(1720)$ and $N(2190)$ to the second and fourth
candidate regions in Bonn--Gatchina also shows that the retained
memberships are not unique boundaries defined by the plots. Keeping
them fixed makes the numerical comparison explicit, but does not
establish an exhaustive partition of the spectrum.

The empirical result is consequently more specific than a claim that
all four groups recur unchanged. Nearby poles assigned to different
partial waves are already present within the Bonn--Gatchina framework.
Such proximity therefore does not require combining the results of
different analysis groups into the PDG compilation. This single example,
however, does not establish that the same pattern occurs in every
model.

The comparison identifies similarities and differences between the
reported pole distributions. It does not determine whether the members
of a candidate group represent independent physical resonances or
reconstructed contributions from a common underlying structure.
The interpretation of these patterns, and their possible connection
with the reconstruction ambiguities discussed earlier, is reserved for
the Discussion.
\section{Discussion}
\label{sec:discussion}

The comparison of the PDG 2026 and Bonn--Gatchina 2025 results gives a
modern setting for H\"ohler's observation: poles assigned to different
partial waves can occupy nearby positions in the complex-energy plane.
Their proximity within the Bonn--Gatchina analysis shows that this
feature does not arise solely from combining results of different
analysis groups in the PDG compilation. However, the comparison also
shows that the compactness and membership of the candidate groups are
not equally stable. The figures establish a descriptive pattern; they
do not determine its physical origin or the number of independent
resonances responsible for it.

The central spectroscopic issue is the uncertainty in the angular
momentum assigned through amplitude reconstruction. Because all reaction
observables are bilinear in the amplitudes, even their complete
measurement leaves information undetermined before additional physical
constraints are imposed. An angle-dependent choice of this missing
information can change the partial-wave decomposition without changing
the observables~\citep{Svarc2018}. Additional constraints can restrict the
admissible choices, but selecting one solution within a model does not
by itself establish the uniqueness of its resonance assignments.
Consequently, evidence for a resonant contribution and confidence in
its assigned angular momentum are distinct questions.

The truncation mechanism studied in Ref.~\citep{Svarc2026} provides a
specific connection between this reconstruction problem and the
appearance of related structures in different partial waves. In a fit
to bilinear observables, truncation restricts the interference structure
that can be represented. The retained coefficients are then obtained
through a coupled nonlinear reconstruction and can become effective
coefficients containing contributions from different underlying
angular-momentum sectors. A pole-bearing contribution can therefore
appear in several extracted partial waves. This is the mechanism that
motivates examining a cluster as possible reconstructed images of a
common underlying contribution.

Redistribution of the same pole-bearing contribution among several
effective partial waves leaves its pole position unchanged. This
mechanism therefore produces coincident pole images. If an observed
cluster has this common origin, its finite spread could arise from
additional uncertainties in reconstructing the amplitudes, including
experimental errors and model dependence in implementing unitarity.
The present work does not determine the contributions of these effects.

Unitarity considerations introduce a further issue. The $4\times4$
example in the Appendix demonstrates that different unitary amplitudes
can produce identical separately measurable channel bilinears within
the construction. It is a counterexample to a general claim that
unitarity necessarily makes the inverse reconstruction unique, rather
than a proof of generic behaviour in every reaction. Moreover, this
on-shell construction does not itself establish different resonance
pole positions or angular-momentum assignments. Such consequences
require an energy-dependent analytic realization with the relevant
angular structure. Its role here is to show that uniqueness cannot be
inferred from imposing unitarity alone.

Practical model dependence is distinct from this exact ambiguity.
Different analytic implementations of the same coupled-channel
unitarity condition can produce different amplitudes and slightly
different bilinears whose differences remain unresolved by the
available measurements. These implementations can differ in their
channel content, background terms, and treatment of singularities.
The resulting amplitudes need not contain the same poles, and their
differences need not amount merely to displacements of a fixed pole
set. Differences in experimental input, coverage, normalization, and
uncertainties provide an additional source of variation between
extractions.

These considerations suggest a conditional interpretation of the
observed clusters. Truncation can distribute a common pole-bearing
contribution among effective partial waves, while subsequent
reconstruction choices can alter the extracted pole content and
positions. Within this interpretation, data uncertainties and the
model dependence of analytic unitary descriptions are possible sources
of departure from ideal coincidence. The present analysis does not
calculate their contributions to the observed spread. In particular,
the exact unitary counterexample does not supply such a calculation,
and no alternative coupled-channel fit with fewer independent
resonances has been constructed here.

The uncertainty in angular-momentum assignment must also be stated
precisely. Physical resonances retain definite $J^P$, and exact unitarity
does not mix sectors separated by rotational invariance and parity
conservation. In the $\pi N$ channel, $L$ is the orbital angular momentum
of that channel, related to the resonance's $J^P$ through coupling to
the nucleon spin. The uncertainty concerns which sector is identified
from the reconstructed amplitude. If several extracted poles are
interpreted as images of a common contribution, their labels alone do
not select the quantum numbers of the underlying resonance. A
determination of that assignment requires testing the competing
admissible reconstructions.

A cluster can also contain genuinely distinct resonances with nearby
pole positions. Accordingly, neither proximity nor separation in the
complex-energy plane settles the reliability of an individual
resonance assignment. Clustering identifies groups for which the
necessity of treating every extracted pole as an independent state
deserves particular examination. The restriction to four-star states
makes this question relevant even to the established spectrum, while
the limited sample and descriptive grouping preclude a general
criterion based on cluster membership. The present results therefore
provide no basis for revising individual PDG ratings.

The constructive consequence is a focused question for subsequent
analyses: do the measured observables require all the independent
resonances presently assigned to a candidate cluster, with their
reported angular momenta? Addressing it will require alternative
descriptions that genuinely permit changes in resonance content and
partial-wave assignment while retaining the physical constraints. If
the relevant choices are already fixed by the parametrization or by
imported complex partial-wave amplitudes, agreement with the data
within that restricted construction cannot decide between the
alternatives considered here. A broader analysis of clustering,
including lower-star and isolated states, and explicit tests of
competing coupled-channel descriptions are therefore proposed as
future research.

\section{Conclusions}
\label{sec:conclusions}

Resonance angular-momentum assignments inherit the uncertainties of
amplitude reconstruction. The published truncation mechanism shows how
a pole-bearing contribution can appear in several effective partial
waves. The unitary counterexample demonstrates that imposing
coupled-channel unitarity alone need not make reconstruction from
measurable bilinears unique, while practical unitary descriptions can
differ in pole content despite describing the available measurements
equally well. Confidence in a resonant contribution therefore does not
automatically establish equal confidence in its assigned angular
momentum.

The comparison of the four-star $N^\ast$ pole positions in PDG 2026 and
Bonn--Gatchina 2025 shows that proximity of poles assigned to different
partial waves is also present within a single coupled-channel
framework. It therefore cannot be attributed solely to the PDG
compilation procedure. Although the detailed candidate groupings
change, these patterns motivate investigating a common origin for some
extracted contributions. They do not demonstrate that the members of
any particular cluster represent one physical resonance.

The present work therefore formulates an unresolved spectroscopic
question and provides a basis for further investigation. Whether the
data require separate states within each candidate cluster, and how
securely their angular momenta can be assigned, remain to be established
through explicit comparisons of competing admissible reconstructions.

Unfortunately, this analysis goes exactly in the opposite direction of the motto of the 21st century: "The quest for new resonances," which was triggered by the abundance of QCD generated resonant states. Instead of increasing it, we tend to reduce it. But, this is the finding of this paper. 
\section{Future research}
\label{sec:future-research}

A natural next step is to extend the clustering analysis to lower-star
resonances for which pole estimates are available, and to examine
apparently isolated states alongside the candidate clusters. A
quantitative assessment should account for pole uncertainties,
correlations where available, and the increasing density of reported
states. This would help establish which groupings remain significant
as the spectrum is enlarged and which depend strongly on the selected
sample. Cluster membership should serve as a guide for further tests
of resonance interpretation.

A complementary investigation should test alternative resonance
contents in contemporary coupled-channel analyses. The same
experimental observables should be fitted with different numbers of
independent resonances, including reduced descriptions of selected
clusters and alternative allowed partial-wave assignments. Each trial
resonance must retain definite $J^P$, while backgrounds and channel
couplings are readjusted consistently. The comparison should examine
the poles of the resulting amplitudes and their assignments, since
counting the resonance terms inserted into a parametrization is not
sufficient to establish its physical pole content. This direction is already under consideration by all groups as a part of the resonance search within one model, but now an additional angle should be added — the uncertainty of the angular momentum quantum number. 

A crucial requirement is sufficient flexibility to represent the
competing interpretations. Sensitivity to assumptions about analytic
backgrounds, channel couplings, and relations between diagonal and
off-diagonal matrix elements should be investigated systematically.
Constraints imported from previously reconstructed complex partial
waves should be varied or relaxed when they already fix the
interpretation under examination. Otherwise, the exercise risks
becoming a consistency check of a chosen representation. Comparisons
across different analytic implementations of the same unitarity
condition would help assess how strongly the conclusions depend on
that representation.

The goal is to distinguish evidence for resonant behaviour from evidence
for the number of independent states and their angular-momentum
assignments. Where competing descriptions predict different
observables, they can guide new measurements. Ultimately, a validated
assessment of stability under admissible reconstruction changes could
supplement existing resonance evaluations, including PDG ratings.
Clustering would identify questions to investigate; confidence in
individual assignments would depend on the outcome of those tests.

\section*{Acknowledgments} 

I acknowledge the use of the Overleaf  AI language model to edit the wording of the text. I further acknowledge the use of the large language model ChatGPT GPT-6.1 Sol Max to evaluate the logical soundness of the presentation and to assess the completeness of the referenced material. However, I declare that I am responsible for all physics statements and conclusions  expressed in this paper, and that they are my own personal creation.

\clearpage

\appendix
\section{An exact continuous ambiguity in four-channel reconstruction}
\label{app:four-channel-ambiguity}
\label{app:four-channel-ambiguity}

Consider four distinguishable, open, spinless two-body channels at a fixed energy, with only the \(S\) wave contributing. We hold the physical channel basis and kinematic factors fixed and use flux-normalized amplitudes,
\begin{equation}
S=I+2iT.
\label{eq:four-channel-normalization}
\end{equation}
Coupled-channel unitarity then requires
\begin{equation}
S^\dagger S=I,
\qquad
\frac{T-T^\dagger}{2i}=T^\dagger T.
\label{eq:four-channel-unitarity}
\end{equation}

Each reaction has one complex amplitude, so its separately measurable bilinear is \(|T_{ij}|^2\), multiplied by known kinematic factors.

Define the continuous family
\begin{equation}
\boxed{
S(\phi)=\frac12
\begin{pmatrix}
1&z&1&z\\
z&1&-z&-1\\
1&-z&1&-z\\
z&-1&-z&1
\end{pmatrix},
\qquad z=e^{i\phi},\quad \phi\in\mathbb{R}.
}
\label{eq:four-channel-S}
\end{equation}

Every row has unit norm, and distinct rows are orthogonal whenever \(|z|=1\). Therefore,
\begin{equation}
S^\dagger(\phi)S(\phi)=I
\end{equation}
for every \(\phi\). The matrices are also symmetric,
\begin{equation}
S^T(\phi)=S(\phi),
\end{equation}
consistent with time-reversal invariance in the chosen channel convention.

The corresponding reaction matrix is
\begin{equation}
\boxed{
T(\phi)=\frac{S(\phi)-I}{2i}
=\frac{1}{4i}
\begin{pmatrix}
-1&z&1&z\\
z&-1&-z&-1\\
1&-z&-1&-z\\
z&-1&-z&-1
\end{pmatrix}.
}
\label{eq:four-channel-T}
\end{equation}

Its diagonal elements are
\begin{equation}
T_{ii}=\frac{i}{4},
\end{equation}
while every off-diagonal element has modulus \(1/4\). Consequently,
\begin{equation}
\boxed{
|T_{ij}(\phi)|^2=\frac{1}{16}
\quad\text{for all }i,j=1,\ldots,4
\quad\text{and all }\phi.
}
\label{eq:four-channel-bilinears}
\end{equation}

Thus, all sixteen separately measurable reaction bilinears---including the elastic ones---are independent of \(\phi\). The diagonal unitarity relations also remain unchanged:
\begin{equation}
\operatorname{Im}T_{ii}
=\sum_{j=1}^{4}|T_{ji}|^2
=\frac14.
\end{equation}

Nevertheless, the amplitudes differ. Their \(S\)-matrix eigenvalues are
\begin{equation}
\operatorname{spec}S(\phi)
=\left\{1,1,e^{i\phi},-e^{i\phi}\right\}.
\label{eq:four-channel-spectrum}
\end{equation}
For distinct values of \(\phi\) within \(0<\phi<\pi/2\), these eigenvalue sets differ. Ordinary channel rephasings,
\begin{equation}
S\longrightarrow D^\dagger SD,
\qquad
D=\operatorname{diag}
\left(e^{i\alpha_1},\ldots,e^{i\alpha_4}\right),
\end{equation}
preserve eigenvalues. The alternatives therefore cannot be identified as changes of channel phase convention.

Interference products involving different reactions can distinguish them. For example,
\begin{equation}
T_{21}(\phi)T_{31}^{*}(\phi)
=\frac{e^{i\phi}}{16}.
\label{eq:four-channel-cross-interference}
\end{equation}
This product varies with \(\phi\), but it is absent from measurements identifying final channels 2 and 3 separately, because those measurements contain no interference between the orthogonal final channels.

The construction therefore exhibits a continuous family of distinct, exactly unitary amplitudes with identical complete sets of separately measurable reaction bilinears in this spinless \(S\)-wave setting. It establishes fixed-energy reconstruction ambiguity; it does not assume or demonstrate different analytically continued pole positions.

\end{document}